\documentclass[11pt,english]{article}
\usepackage[utf8]{inputenc}
\usepackage[margin=2cm,includefoot]{geometry}
\usepackage{fancyhdr}
\usepackage{units}
\usepackage{amsmath}
\usepackage{amsthm}
\usepackage{stackrel}
\usepackage{esint}
\makeatletter

\numberwithin{equation}{section}
\numberwithin{figure}{section}

\makeatother

\usepackage{babel}

\begin{document}
	
	\title{\textbf{On the Origin of Inertia: Implications for Dark Matter and Dark Energy.    }}

	\author{ \textbf{ Konstantinos I. Tsarouchas }}

	\date{}
	
\maketitle

\begin{center}

	{\small{{Department of physics,  National Technical University of Athens, Greece }}}\\
	{\small{{ }}}
	
	E-mail-1: ko.tsarouchas@gmail.com - E-mail-2: ktsarouc@central.ntua.gr 
\end{center}

\maketitle

\begin{abstract}

	In this paper, we present a new theory explaining the origin of inertia based on two key ideas: gravity as a spin-1 gauge field theory and the relativity of all kinds of motion. This theory proposes that inertial mass is influenced by the distribution of matter across the Universe, offering potential insights into dark matter and dark energy. For gravity to be described by a spin-1 gauge field theory, we propose that gravitational mass $m$, distinct from inertial mass, is a Lorentz invariant and should be replaced by an imaginary mass $im$ for like masses to attract. According to this theory, while gravitational mass is imaginary, inertial mass remains a real quantity. These two key ideas, lead to the principle of Equivalence and the conclusion that gravity shapes the geometry of spacetime, which is Finsler-Randers spacetime. However, for bodies with gravitational mass, this curved spacetime is equivalent to a flat Minkowski spacetime with an additional gravitomagnetic field. Using this, the theory shows that external inertial forces  are inductive effects from the entire Universe, while internal forces depend on the body’s structure. In free fall, all bodies experience only external forces, explaining why they fall at the same rate regardless of internal structure.

\end{abstract}


\section{Introduction}
\label{intro}
The origin of inertial forces is a problem which has been of great
concern to many thinkers since the time of Newton, but which so far
has escaped a satisfactory solution. This leaves room for a new 
attempt. The inertial forces, which arise in non-inertial (accelerating) frames of reference, raise the fundamental question:
What determines an inertial frame? 

The first answer comes from Descartes and Newton \cite{key-1}, according to which,
an inertial frame of reference is a frame that moves with constant velocity,
with respect to the absolute space and the motion is absolute. The
inertial forces, such as the centrifugal force, must arise from acceleration
with respect to the absolute space. This idea implies that space is
an absolute physical structure with properties of its own and the
inertia is an intrinsic property of the matter.

The second answer comes from Leibniz, Berkeley and Mach and is known
as Mach’ principle, according to which, an inertial frame of reference
is a frame that moves with constant velocity, with respect to the
rest of the matter in the Universe, and the motion is relative. The
inertial forces, such as the centrifugal force, are more likely caused
by acceleration, with respect to the fixed stars  \cite{key-2}\cite{key-3}\cite{key-4}.
This idea implies that the properties of space arise from the matter
contained therein and are meaningless in empty space. 

The distinction between Newton’s and Mach’s considerations, is not
one of metaphysics but of physics, for if Mach were right then a large
mass could produce small changes in the inertial forces observed in
its vicinity, whereas if Newton were right then no such effect could
occur \cite{key-5}.  This debate has important implications for  issues such as dark matter  \cite{key-6}\cite{key-7}\cite{key-8}\cite{key-9}\cite{key-10}  \cite{key-11} and dark energy  \cite{key-12}  \cite{key-13} \cite{key-14}  \cite{key-15} \cite{key-16}.

The idea that the only meaningful motion of a particle, is motion relative
to other matter in the Universe, has never found its complete expression
in a physical theory. The Special theory of Relativity eliminated
absolute rest from physics, but acceleration remains absolute in this
theory. Albert Einstein was inspired by Mach's principle. The General
theory of Relativity, attempted to continue this relativization and
interpret inertia considering that it is the gravitational effect
of the whole Universe, but as pointed out by Einstein, it failed to
do so. Einstein showed that the gravitational field equations of General
Relativity imply that a body, in an empty Universe, has inertial properties
\cite{key-17}\cite{key-18}\cite{key-19}.

Despite its profound success in describing many aspects of gravity and spacetime, General Relativity has significant limitations. It fails to offer a satisfactory explanation for the fundamental nature of inertia, as well as  of dark matter and dark energy. Additionally, the long-standing quest for a quantum theory of gravity remains unresolved, leaving a significant gap in the effort to unify General Relativity with quantum mechanics.
The principle of Equivalence is an essential part of General Relativity.
But although the principle of Equivalence has been confirmed experimentally
to high precision, the gravitational field equations of General Relativity
have not as yet been tested so decisively. Thus, it is not a theory
fully confirmed experimentally and competing theories cannot be ruled
out \cite{key-20}. 

In this paper, we propose a new approach to understanding inertia by embracing two key ideas. In Section 2, we argue that gravity is described by a spin-1 gauge field theory, with gravitational mass, distinct from inertial mass, as a Lorentz invariant. In Section 3, we explore the idea of the relativity of all kinds of motion, leading to the principle of equivalence and the conclusion that gravity curves spacetime which is a Finsler-Randers spacetime. However, for a body with gravitational mass, a curved Finsler–Randers spacetime is physically equivalent to a flat Minkowski spacetime with an additional gravitomagnetic field. Using this in Section 4, we develop a theory consistent with Mach's principle that provides an explanation for inertia. To ensure that like masses attract in a spin-1 gauge field theory, we replace gravitational mass $m$ with an imaginary mass $im$, while inertial mass remains real \cite{key-21}. Finally, in Section 5, we discuss how the theory’s proposal that inertial mass is shaped by the distribution of matter in the Universe impacts our understanding of Dark Matter and Dark Energy.

\section{Gravitomagnetic theory }
\label{sec1}
The first step in explaining the origin of inertia and the nature of inertial mass is to recognize that gravity should be described by a gravitomagnetic theory just like the electromagnetic theory, where gravitational mass is a Lorentz-invariant quantity and distinct from inertial mass. Nevertheless, we will demonstrate that all bodies, regardless of their composition, fall at the same rate in a gravitational field. Thus, in this paper we will use the gravitational mass  $m$, the inertial mass $ m_{in}$ and the inertial rest mass $ m_{in-0}$.

According to Richard  Feynman, we can reconstruct the complete electrodynamics
using  the Lorentz  transformations (for coordinates,  velocities, potentials,  forces)  
and the following series of remarks
\cite{key-22} \cite{key-23}:
\begin{enumerate}
	\item The Coulomb potential at a  distance $r$ from  a stationary  point-charge $q$ in vacuum is: $\varphi_{e}=\dfrac{1}{4\pi\varepsilon_{0}}\dfrac{q}{r}$
	\item An electric point-charge produces a scalar potential $\varphi_{e}$ and a vector potential $ \vec{A}_{e}$,
	which together form a four-vector, $A_{e}=\left(\dfrac{\varphi_{e}}{c},\vec{A}_{e} \right)$
	\item The potentials produced by a point-charge moving in any way, depend only
	upon the velocity and position at the retarded time.
\end{enumerate}
where $ \varepsilon_{0}$ is the vacuum permittivity and  $c$  the speed of light in vacuum. 
Of course we need to know how to get the Coulomb's law from the scalar potential.

Therefore, if we want to obtain a gravitomagnetic theory, with equations that have the same
	mathematical form, as those of the
	electromagnetic theory, first we consider that the gravitational mass is a Lorentz invariant and second that the same series of remarks
	must be met for gravity. We  already have  the first remark, that is, the gravitational  potential at a  distance $r$ from a stationary gravitational point-mass  $m$ in vacuum is, 
\begin{equation}
\varphi_{g}=-G\dfrac{m}{r}
\end{equation}   
    where 
$G$ is the  gravitational constant, but this is
only the one remark. 
Therefore, we need the other two, as well.   We
will obtain them  with the following  two principles:
\paragraph*{Principle 1}
\begin{quote}
	\textbf{A gravitational point-mass produces a scalar potential $\varphi_{g}$ and a vector
		potential $ \vec{A}_{g}$, which together form a four-vector, $A_{g}=\left(\dfrac{\varphi_{g}}{c},\vec{A}_{g}\right)$}
\end{quote}
\paragraph*{Principle 2}
\begin{quote}
	\textbf{The potentials produced by a gravitational point-mass moving in
		any way, depend only upon the velocity and position at the retarded
		time.}
\end{quote}

So, the potentials produced by a gravitational point-mass m moving with any velocity 
have the same mathematical form  as the Lienard-Wiechert potentials
for an electric point-charge moving with any velocity, but with a negative  sign,
\begin{equation}
\begin{array}{ccccc}
\varphi_{g}=-G\left[\dfrac{m}{r-\vec{r}\vec{v}/c}\right]  &  and  &  \vec{A}_{g}=-\dfrac{G}{c^{2}}\left[\dfrac{m\vec{v}}{r-\vec{r}\vec{v}/c}\right]=\dfrac{1}{c^{2}}\left[\varphi_{g}\vec{v}\right]
\end{array}
\end{equation}
where $\vec{r}$ is the vector from the  gravitational point-mass to the point where the potential is evaluated    and the quantities $r$, $\vec{r}$ and $ \vec{v}$ (the velocity of the point-mass)  in the square  bracket are to have their values at the retarded time. 
From these potentials, we can derive the fields using the following equations
\begin{equation}
\vec{E}_{g}=-\vec{\nabla}\varphi_{g}-\dfrac{\partial\vec{A}_{g}}{\partial t}
\end{equation}
\begin{equation}
\vec{B}_{g}=\vec{\nabla}\times\vec{A}_{g}
\end{equation}
When a gravitational mass $m$ moves with velocity $\vec{v}$ in the above fields, it experiences the gravitomagnetic Lorentz force,
\begin{equation}
\vec{F}_{g}=m(\vec{E}_{g}+\vec{v}\times\vec{B}_{g})
\end{equation}
where $\vec{E}_{g}$ is the gravitational field and $\vec{B}_{g}$
the gravitomagnetic field. 

Therefore, we expect the existence of gravitomagnetic radiation propagating
in vacuum at the speed of light but with a significant difference compared to electromagnetic radiation.
It is well known that an isolated electric source can radiate electric dipole radiation,
with power proportional to the square of the second time derivative
of the electric dipole moment. 
However, an isolated gravitational
source,  cannot radiate gravitational dipole radiation, but quadrupole
and radiation of higher polarity. The reason is simple. The electric
dipole moment can move around with respect to the center of the inertial mass but
the gravitational dipole moment   is identical in location with the
center off the inertial mass, and due to the law of conservation of momentum,   cannot
accelerate or radiate \cite{key-24}.

The gravitomagnetic theory must be described in flat Minkowski space-time by a spin-1 gauge field theory, because it is just like the electromagnetic theory. We will show that, the well-established belief that we cannot have a spin-1 gauge  field theory of gravity,  may be wrong.

The free Dirac Lagrangian  of a particle is
\begin{equation}
\mathcal{L}=i\hbar c\bar{\Psi}\gamma^{\mu} \partial_{\mu}\Psi-m_{in-0}c^2\bar{\Psi}\Psi
\end{equation}
where $ m_{in-0}$ the inertial rest mass of the particle, which we will prove later that is a scalar. It is well known that the free Dirac Lagrangian   is invariant under the global gauge transformation  \cite{key-25}
\begin{equation}
\Psi \rightarrow e^{i\theta} \Psi 
\end{equation}
where $ \theta$ is a real number. But the Lagrangian   is not invariant
under the local gauge transformations
\begin{equation}
\Psi \rightarrow e^{i\theta (x)} \Psi 
\end{equation} 
where $ \theta(x)$ is now  function of  $ x^i$. Under the local gauge transformation we get
\begin{equation}
\mathcal{L'}=\mathcal{L}-  \hbar c (\partial_{\mu}\theta )\bar{\Psi} \gamma^{\mu}  \Psi  
\end{equation}
If we define
\begin{equation}
\theta(x)= - \dfrac{q}{\hbar c}  \lambda_{e} (x) -\dfrac m{\hbar c}  \lambda_{g} (x)  
\end{equation}
where $q$  the electric charge and  $m$  the gravitational mass of the particle, the equation (2.11) becomes 
\begin{equation}
\mathcal{L'}=\mathcal{L}+ [   q (\partial_{\mu}\lambda_{e}) +m  (\partial_{\mu}\lambda_{g})] \bar{\Psi} \gamma^{\mu}  \Psi  
\end{equation}
Now, if we demand that the complete Lagrangian must be 
invariant under local gauge transformation, we are forced to add something to soak up
the extra term
\begin{equation}
\mathcal{L}=[i\hbar c \bar{\Psi}\gamma^{\mu} \partial_{\mu}\Psi-m_{in-0}c^2\bar{\Psi}\Psi]-(\bar{\Psi} \gamma^{\mu}  \Psi) [qA_{e} +m A_{g}]
\end{equation}
where the vectors $  qA_{e}$ and $  mA_{g}$ are transformed under the local gauge transformation according to the rule
\begin{equation}
\begin{array}{ccccc}
qA_{e}\rightarrow  qA_{e} +q\partial_{\mu}\lambda_{e} &   and      & m A_{g}\rightarrow m A_{g} +m\partial_{\mu}\lambda_{g}
\end{array}
\end{equation}
The full Lagrangian must include the free 
terms for the gauge fields.    Thus, the full Lagrangian becomes 
\begin{equation}
\mathcal{L}=[i\hbar c \bar{\Psi}\gamma^{\mu} \partial_{\mu}\Psi-m_{in-0}c^2\bar{\Psi}\Psi]-(\bar{\Psi} \gamma^{\mu}  \Psi)(qA_{e}  +   m A_{g} )
-\dfrac{1}{16\pi }  F^{\mu\nu} F_{\mu\nu} -\dfrac{1}{16\pi }  G^{\mu\nu} G_{\mu\nu} 
\end{equation}
where
\begin{equation}
\begin{array}{ccccc}
F^{\mu\nu}\equiv\partial^{\mu}A^{\nu}_{e}-\partial^{\nu}A^{\mu}_{e} &   and      &  G^{\mu\nu}\equiv\partial^{\mu}A^{\nu}_{g}-\partial^{\nu}A^{\mu}_{g} 
\end{array}
\end{equation}
the electromagnetic and gravitomagnetic tensor respectively.
The full Lagrangian  is now locally gauge 
invariant,  by introducing  the electromagnetic field $ A_{e}$ and the  gravitomagnetic field $ A_{g}$. Both fields must be mass-less, otherwise the invariance will be lost.

However, as Richard Feynman  writes  in “Lectures on Gravitation” \cite{key-26}: 
“A spin-1 theory would be essentially the same as electrodynamics. There is nothing to forbid the existence of two spin-1 fields, but gravity can't be one of them, because one consequence of the spin-1  is that likes repel, and unlikes attract.”  The well-established belief that gravity cannot be described by a spin-1 gauge field theory compels us to confront this issue before fully understanding the nature of inertial rest mass. To address this, we propose replacing the gravitational mass $m$  with $im$ (where $i$ is the imaginary unit). This substitution is justified because, as demonstrated in equations (4.24) and (4.25), the inertial rest mass is roughly defined as the product of two gravitational masses. Consequently, every measurable quantity, such as inertial mass and kinetic energy, remains a real quantity
 	while the gravitational mass itself is not a measurable quantity. In  any Feynman diagram describing the gravitational interaction, the gravitational mass will appear as a square via the coupling constant, since always two vertices are involved. Thus, by replacing the $m$  with $im$, it will change the sign of the energy  corresponding to this diagram so that likes attract   in the gravitomagnetic theory. 
However, for simplicity, we will continue to express gravitational mass as $m$ (a real quantity) in the subsequent discussion.

\section{General relativity  of  motion}
\label{sec2}

According to Richard Tolman, the fundamental principles of Einstein's General theory of relativity, namely the principle of  General Covariance and the principle of Equivalence, may be regarded as based on the fundamental idea of the relativity of all kinds of  motion \cite{key-27}.
We will now follow this fundamental idea. In accordance with this idea,    we can detect
and measure the motion  of a given body, relative to other
bodies, but cannot assign any meaning to its absolute motion. The Special theory of relativity makes only a restricted use of this general idea, since it merely assumes the relativity of uniform translatoty motion in a region of free space where  gravitational effect can be neglected. Thus, the Special theory of Relativity eliminated the concept of absolute rest from physics, but acceleration remains absolute in this theory.  In accordance with the fundamental idea of the relativity of all kinds of motion, an  observer  inside an  accelerated rocket   cannot distinguish whether  the rocket is  accelerated  and  the  remainder  of the Universe, matter and fields, is at rest or  whether the rocket is at rest and the  remainder  of the Universe, matter and fields, is accelerated in the opposite direction. 

In order to ensure the relativity of all kinds of motion  the laws
of physics should have the same mathematical form in all  frames of reference 
since otherwise the difference
in form  could provide a criterion
for judging the absolute motion.
So, we accept the next principle: 

\paragraph*{Principle 3 - The principle of  General Covariance}
\begin{quote}
	\textbf{The laws of physics have the same mathematical form in all frames of reference. }
\end{quote}
That is, in tensor form.  In inertial frames of reference the laws of physics  reduce to  simpler mathematical forms that  agree with the laws of  Special theory of relativity.

The fact that the expression of the equations of physics in a form which is independent of the motion of a reference frame relative to the fixed stars, does not in general prevent a change in their numerical content when we change from one reference frame to another.
However, having a gravitomagnetic theory just like the electromagnetic theory, we have the law of induction given by equation (2.3). This suggests that an induced gravitational field should appear in a reference frame that is accelerated relative to the fixed stars.  Thus, by relating  the changes in numerical content, when we change from one reference frame to another, with  changes in the induced gravitational  field,   we are able to eliminate the criteria for absolute motion and to preserve the idea of the relativity of all kinds of motion.
Therefore, we accept the next principle: 

\paragraph*{Principle 4 - The  principle of Equivalence }
\begin{quote}
	\textbf{Physics in a non
		accelerating   frame S, with a  uniform gravitational field where
		all the released bodies fall with  acceleration $\vec{g}$, is equivalent to physics in a local frame without gravity  but with  translational acceleration  $ \vec{a}=-\vec{g}$ and velocity zero with respect 
		to the inertial frame in which the non accelerating frame S is at rest. }
	
	or, 
	
	\textbf{Physics in a local frame freely falling in a gravitational field is equivalent to physics in an inertial frame without gravity.}  
\end{quote}

In this section, we rely on the well-established experimental fact that all bodies experience the same acceleration in a gravitational field. This implies that the ratio of gravitational mass to inertial rest mass is constant for all freely moving bodies in a gravitational field and can be assumed, for now, to be equal to one. In the following section, we will explain why this happens by examining the induced gravitational field and the nature of inertial rest mass in greater detail.

Using  the Special theory of relativity we are able to describe what physical effects are observed by an observer at rest in a uniformly accelerated frame of reference.
The most well-known of them, apart from the inertial forces, are  \cite{key-28}: 
\begin{enumerate}
	\item Redshift or blueshift of a light ray moving parallel to the direction
	of the acceleration.
	\item Varying coordinate speed of light; fixed local relative speed of light. 
	\item Space-time is endowed with a metric. 
	\item Maximum proper time as the law of motion of freely moving bodies. 
    \item Horizon
	
\end{enumerate}
According to the   principle of Equivalence the same effects must occur in a gravitational field. Therefore, the space-time is endowed
with a metric and the gravitational field affects the space-time metric
so that, the maximum proper time is the law of motion of a freely moving
body in a gravitational field. The two above physical effects  are
so important that we will elevate them to physical principles:    
\paragraph*{Principle 5 - The Principle of Space-time Metric}
\begin{quote}
	\textbf{ The space-time interval
		between two neighbouring points events is:}
		\textbf{
		\[
		ds^2=g_{\mu\nu} (x^i, dx^i ) dx^{\mu}dx^{\nu}
		\]
	}
	\textbf{where ${g_{\mu\nu}} (x^i, dx^i )$ the metric tensor which depends  not only on the position but also on the direction/velocity (Finsler geometry).}
\end{quote}
\paragraph*{Principle 6 - The Principle of Geodesic Motion or of Maximum Proper Time}
\begin{quote}
	\textbf{A freely moving body  always moves along a geodesic: 	$\delta\int ds=0$ }
\end{quote}
Therefore, the fundamental idea of the relativity of all kinds of motion leads us to the conclusion that the gravitomagnetic field affects the  geometry of space-time. However, the  space-time now is not a pseudo-Riemannian space-time but a Finsler-Randers space-time \cite{key-29}. 

Let us now see how the gravitomagnetic field affects the  space-time.  We will follow the Randers approach  where the equation of motion of a test-body in a gravitomagnetic field results naturally as the geodesic of a Finsler-Randers space-time \cite{key-30} \cite{key-31}. 
 The lagrangian of a test-body of gravitational mass $m$  and inertial rest mass $m_{in-0}$ (which we will prove later that is a scalar)   moving with four-velocity $\dot {x^{\mu}}$ in a gravitomagnetic field is 
 \begin{equation}
 L= \sqrt {n_{\mu\nu} (x)  \dot {x^{\mu}}\dot {x^{\nu}}} +\dfrac {m}{m_{in-0}c^2} { A_{g-\mu}}\dot {x^{\mu}}
 \end{equation} 
 where $ {n_{\mu\nu}} \equiv diag(1,-1,-1,-1)$ the Minkowski metric, $A_{g-\mu}$  the gravitational four-potential and $\dot {x^{\mu}}= dx/d\tau$ the four-velocity where $ d\tau= \sqrt {n_{\mu\nu} (x)  \dot {x^{\mu}}\dot {x^{\nu}}} $. The first variation of the action coresponding to the langrangian (3.1) gives the Euler-Lagrange equations
\begin{equation}
\dfrac {d} {d\tau}(\dfrac {\partial L} {\partial \dot {x^{\mu}} }  )-\dfrac {\partial L} {\partial x^{\mu}}=0
\end{equation}  
If we substitute the explicit form of the Lagrangian (3.1) in (3.2) we get the gravitomagnetic Lorentz equation of motion 
 \begin{equation}
\dfrac {d^2x^{\mu}}{d \tau^2}=\dfrac {m}{m_{in-0}c^2}G^{\mu  \nu} \dfrac{dx_{\nu}}{d \tau } 
\end{equation}  
where $G^{\mu\nu}\equiv\partial^{\mu}A^{\nu}_{g}-\partial^{\nu}A^{\mu}_{g}$, the gravitomagnetic-field tensor.

In order for the  gravitomagnetic field to affect the  space-time, we accept   a Finsler-Randers space-time and  we identify  the metric function $F(x, \dot {x}) $ of this space-time with the Lagrangian (3.1). So we get a Finsler-Randers space-time  with  metric function given by the following principle: 
\paragraph*{Principle 7 - The metric function of the Finsler-Randers spacetime is: }
\begin{quote}
	 \begin{center} 
		
		$  F(x, \dot {x})=\sqrt {n_{\mu\nu} (x)  \dot {x^{\mu}}\dot {x^{\nu}}} +\dfrac {m}{m_{in-0}c^2} { A_{g-\mu}}\dot {x^{\mu}} $
	\end{center}
\end{quote}
In this case  the four velocity is $ \dot {x^{\mu}}= dx/ds$, where $ ds$ is the Finsler-Randers proper time, because the measurable quantity in Finsler-Randers space-time is $ds$  and not $d\tau$. In the absence of gravity, $ds= d\tau$. So, whenever exist a gravitomagnetic field in a region of space-time the space-time becomes Finslerian and the isotropy breaks.
The metric function represent the distance $ds$ between two neighbouring points represented by the coordinates $x^i$ and $x^i+dx^i$
 \begin{equation}
ds= F(x^i,  dx^i)
\end{equation}
  Then, using the principle of geodesic motion,  the equation of motion of a test-body in a gravitomagnetic field, i.e. the gravitomagnetic Lorentz force, follows  as the geodesic of  this Finsler-Randers space-time 
  \begin{equation}
  \dfrac {d^2x^{\mu}}{d s^2}=\dfrac {m}{m_{in-0}c^2}G^{\mu  \nu} \dfrac{dx_{\nu}}{d s } 
  \end{equation}
 As previously mentioned, we assume that the ratio of gravitational mass to inertial rest mass is equal to one for all freely moving bodies in a gravitational field. Consequently, 
  the equation of motion of a body follows naturally from the Finsler-Randers space-time geometry and is the same for all bodies.

   A particle with only gravitational mass in a Finsler-Randers space-time,   moving along a geodesic, obeys the gravitomagnetic Lorentz equation. Thus, the motion of such a particle in a gravitomagnetic field can be described either 
   by using forces in Minkowski space-time, or by stating that the gravitomagnetic field curves spacetime, causing the freely moving particle to follow a geodesic. For a body with gravitational mass, the concept of a curved Finsler-Randers space-time
   is physically equivalent to the concept of a  flat Minkowski space-time with an additional gravitomagnetic field.  It is required to accept that space-time is curved only when we consider the motion of electromagnetic and gravitomagnetic waves or particles whithout gravitational mass.

   The metric tensor of a Finsler-Randers space-time is given by the equation
   \begin{equation}
   g_{\mu\nu} (x^i,\dot x^i ) = \dfrac {1}{2}  \dfrac {\partial^2 F^2 (x^i,\dot x^i )} {\partial \dot x^{\mu} \partial \dot x^{\nu}  }
   \end{equation}
   and it depends not only on the position but also on the velocity of the test-body.
   The metric function of the space-time may
   be given in terms of $ g_{\mu\nu}$ as
   \begin{equation}
   F^2 (x^i,\dot x^i )=g_{\mu\nu} (x^i,\dot x^i ) \dot x^{\mu}\dot x^{\nu} 
   \end{equation}  
   or, using the equation (3.4): $ds= F(x^i,  dx^i) $, in terms of differentials  
   \begin{equation}
   ds^2=g_{\mu\nu} (x^i, dx^i ) dx^{\mu}dx^{\nu}
   \end{equation}
  According to P. Stavrinos \cite{key-32},  when the speed of the test body is zero,  $\dot x^{\mu}= (c,0,0,0)$,  we get from the metric tensor  
  \begin{equation}
  g_{0 0}  = 1+  \kappa^2\varphi_{g}^2+ 2\kappa \varphi_{g} 
  \end{equation}
  where $ \kappa=\dfrac {m}{m_{in-0}c^2} $ .
  Since the second term is negligible compared to the third, we obtain
  \begin{equation}
  g_{0 0} \approx 1+ 2\kappa \varphi_{g} = 1+ 2\dfrac {m}{m_{in-0}c^2} \varphi_{g}
  \end{equation}
  Therefore,   in spherical coordinates, outside and at a distance $r$  from the center of a static and stationary
  body  with spherically symmetric distribution of gravitational mass M, the equation (3.10) becomes
  \begin{equation}
  g_{0 0} =\left(1-\dfrac{m}{m_{in-0}}\dfrac{2GM}{c^{2} r}\right)
  \end{equation}  
 The equation (3.11) represents the metric coefficient for time in spacetime, and this coefficient directly affects the rate at which time passes for a stationary observer in a gravitational field, leading to gravitational time dilation and also to  gravitational redshift of a light wave as it moves upwards against a gravitational field. If the
 ratio of the gravitational mass to the inertial rest mass is equal to one, for the above phenomena we
 get the same results as the General theory of relativity.

  Since, for a body with gravitational mass, the concept of curved Finsler-Randers spacetime is physically equivalent to  the concept of a 
 flat Minkowski spacetime with an additional gravitomagnetic field, we will proceed  in the next section by working with forces in Minkowski spacetime.

\section{Inertia}
\label{sec3}

\subsection{Gravitational inertial rest mass of a body without internal structure}
\label{sec4}
Let’s now explore how to determine the gravitational field experienced by an accelerating body.
In accordance with the fundamental idea of the relativity of all kinds of motion, an observer inside an accelerated rocket cannot distinguish whether the rocket is accelerating while the fixed stars and their fields are at rest, or whether the rocket is at rest while the fixed stars and their fields are accelerating in the opposite direction. Therefore, for an accelerating observer  the fields of the distant stars are "dragged along" with them, just like the fields of stars that move with uniform velocity. Thus,  there is no radiation field from the fixed stars for the accelerating observer. The fields  of the fixed stars for the accelerating observer, are similar to what they would be if those stars were moving in a straight line at a constant speed.  So, we conclude that we can determine the instantaneous gravitational  potentials of the fixed stars for an accelerating observer at any given moment, by applying the Lorentz transformations, using as velocity the instantaneous velocity of the observer  relative to the fixed stars   at that same moment.

Let us now apply the previous conclusion to a thought experiment, the lab frame experiment, in order to determine the inertial force.   We suppose that we use a space station, which is far from any massive  body, as a laboratory.  We will call the local inertial  frame where the space station is always at rest, the lab frame.
The lab frame, as a local inertial frame, is only expected to function over a small region of space.  We assume that the distribution of matter in the Universe is such that the gravitational field in the lab frame is zero. This means that the gravitational scalar potential  $\varphi_{g}$ of the entire Universe has the
same value everywhere in the lab frame 
\begin{equation}
\vec{\nabla}\varphi_{g}=0
\end{equation}
We also suppose that the Universe expands symmetrically in all directions,
with respect to the lab frame, so that the gravitomagnetic vector
potential due to one part of the mass-current, is canceled out by
the vector potential due to another part of the mass-current, owing
to its symmetry.  Therefore, the gravitomagnetic vector potential
	$\vec{A}_{g}$ from the entire Universe in the
	lab frame is zero, 
\begin{equation}
\vec{A}_{g}=0
\end{equation}
This would also happen if all the bodies of the Universe were at rest,
relative to the lab frame. So, we can say that the lab frame is at
rest relative to the Universe, or at rest relative to the fixed stars.

 Now, consider a point-particle, the test-body K, initially at rest in the lab frame, which begins to accelerate along the  $x$ axis. 	We will call the local proper frame of the  test-body K  (the local frame  where the test-body K is always at rest)  the  K frame. The K frame, as a local proper frame, is only expected to function over a small region of space. Based on the previous conclusion,  we can determine the potentials of the fixed stars in the K frame  by applying the Lorentz transformations to the potentials measured in the lab frame. Therefore, when the instantaneous velocity of the
	test-body K  is $v$ in the positive x-direction as measured in the lab frame, the Lorentz transformations   give  the  gravitational scalar  potential $\varphi_{g}'$ and the gravitomagnetic vector potential  $\vec{A}_{g}'$ 
in the K frame  as follows 

\begin{equation}
\begin{array}{ccccc}
\varphi'_{g}=\gamma(v)(\varphi_{g}-vA_{g-x}), &         & A'_{g-y}=A_{g-y}    \\
\\
A'_{g-x}=\gamma(v)(A_{g-x}-\dfrac{v}{c^{2}}\varphi_{g}), &             & A'_{g-z}=A_{g-z}, &   & \gamma(v)=\dfrac{1}{\sqrt{1-\dfrac{v^{2}}{c^{2}}}}
\end{array}
\end{equation}
 Since $ \vec{A}_{g}=0$, then  ${A}_{g-x}={A}_{g-y}={A}_{g-z}=0 $. Therefore, using vector notation,  the potentials in the	 K frame  are    
\begin{equation}
\varphi'_{g}=\gamma(v)\varphi_{g}
\end{equation}
\begin{equation}
\vec{A'}_{g}=-\dfrac{1}{c^{2}}\gamma(v)\varphi_{g}\vec{v}=-\dfrac{1}{c^{2}}\varphi'_{g}\vec{v}
\end{equation}
As test-body K accelerates, these potentials change, leading to the appearance  of an induced gravitational field according to equation (2.3):
\begin{equation}
\vec{E'}_{g}=- \vec{\nabla'}\varphi'_{g}  -\dfrac{\partial\vec{A}_{g}'}{\partial t'}
\end{equation}
where $\partial t'$ is the time interval in the K frame.
The gravitomagnetic field $\vec{B'}_{g} $ in  K frame is zero because the fixed stars undergo translatory motion relative to test-body K:
\begin{equation}
\vec{B'}_{g}=  \vec{\nabla'}\times\vec{A'}_{g}=0
\end{equation}
Given that $\gamma(v)$
is uniform across the  K frame,  the scalar potential $\varphi'_{g}$ is also uniform, implying 
\begin{equation}
\vec{\nabla'}\varphi'_{g}=0
\end{equation}
Thus, the gravitational field in  K frame simplifies to 
\begin{equation}
\vec{E}'_{g}=-\dfrac{\partial\vec{A}_{g}'}{\partial t'}
\end{equation} 
If test-body K possesses a gravitational mass $m$, it will experience an induced gravitational force
\begin{equation}
\vec{F'}_{g}=m\vec{E'}_{g}=-\dfrac{\partial\left(m\vec{A}'_{g}\right)}{\partial t'}
\end{equation}
If we assume now that  the gravitational scalar potential $\varphi_{g}$ is independent of time (that's why we call the stars, fixed stars), 
substituting for $\vec{A'_{g}}$ from equation (4.5) into equation
(4.10), we get
\begin{equation}
\vec{F'}_{g}= -\left(-\dfrac{1}{c^{2}}m\varphi_{g}\right)\dfrac{\partial\left[\gamma(v)\vec{v}\right]}{ \partial t'}=   -\left(-\dfrac{1}{c^{2}}m\varphi_{g}\right)\gamma^3(v)\dfrac{d \vec {v}}{dt'}
\end{equation} 
Given that the gravitational scalar potential is negative, it is evident from equation (4.11) that the induced gravitational force on the test body K opposes any change in its velocity.  It is an inertial force!
 
We will call the inertial force given by equations (4.10),  external gravitational inertial force $\vec{F}'_{inert} $,  because it arises from the body's acceleration relative to the fixed stars.  So,
 \begin{equation}
 \vec{F}'_{inert}=\vec{F'}_{g}= -\dfrac{\partial\left(m\vec{A}'_{g}\right)}{\partial  t'}  
 \end{equation}
Therefore, an inertial reference frame is a frame moving at a constant velocity relative to the fixed stars, while an accelerating reference frame is one accelerating relative to the fixed stars. The only difference between
these two types of frames is the induced gravitational field.  Thus,  an accelerating frame is simply an inertial frame with an additional induced gravitational field.
 
 According to the principle of Equivalence,  physics inside a small elevator in free fall within a gravitational field is equivalent to  physics in an inertial frame, i.e., a frame without a gravitational field. Therefore, for the total gravitational force inside the freely falling elevator to be zero, the inertial force that arises is an induced gravitational force, equal in magnitude but opposite in direction to the gravitational force accelerating the elevator.
 From this, we can conclude that to accelerate a body relative to the fixed stars, an external force must be applied, equal in magnitude but opposite in direction to the inertial force acting on the body. Consequently,  the total force  in the proper frame  of a body (the frame where the body  is always at rest) must be zero, whether the body is moving with uniform velocity or is being accelerated relative to the fixed stars. So, we accept the following law of motion:

 \begin{quote}
 	\textbf{The motion of a body is such that, in its proper frame  the total force on the body is always zero.}  
 \end{quote} 
Therefore, the force that accelerates a body and the inertial force experienced by the body in its proper frame are equal in magnitude but opposite in direction.

In addition to the external inertial force, there is also an internal
	inertial force. This is a well known effect which has the name radiation reaction \cite{key-33} \cite{key-34}. We do not know exactly the mechanism that causes it but we know that it exists. The  picture is something like
this: We can think that a body consists of many particles. When the
body is at rest or it’s moving at uniform velocity, every particle
exerts a force on every other, but the forces all balance in pairs,
so that there is no net force. However, when the body is being accelerated,
the internal forces will no longer be in balance, because of the fact
that the influences take time to go from one particle to another.
With acceleration, if we look at the forces between the various particles
	of the body, action and reaction are not exactly equal, and the body
	exerts a force on itself that tries to hold back the acceleration.  
We will call this self-force  internal inertial force, since it depends on the internal structure of the body.
	 
Since an accelerating frame is  an inertial frame with an induced gravitational field, the self-force is a result of this induced field. However, according to the principle of Equivalence, when a body is in free fall, the total gravitational field in the body's proper frame is zero. As a result, the self-force also vanishes. Therefore, we can conclude that:
\begin{quote}
	\textbf{When a body is in free fall within a gravitational field, the internal structure of the body    plays no role and thus,  only the external gravitational inertial force acts on the body.} 
\end{quote}

We can obtain some very important and useful results using non-relativistic velocities.  
 So,  for   non-relativistic
 velocities,  from equation (4.10), the external gravitational inertial force on the accelerating test-body K is 
\begin{equation}
\vec{F}'_{inert}  =- \left(-\dfrac{1}{c^{2}}m\varphi_{g}\right)\dfrac{d\vec{v}}{dt}=\left(-\dfrac{1}{c^{2}}m\varphi_{g}\right)\left(- \vec{a}\right)
\end{equation}
where $dt$ is the time interval in the lab frame and  $\vec{a}$ is  the acceleration with respect to the lab frame.
Now, let us consider the case where the test-body K has no internal structure. 
 When this body is  accelerated by
a force $\vec{F}$,  it experiences  no internal inertial forces but only the external
gravitational inertial force.   According to the Law of motion,
in the proper frame of the test-body K, the total net force acting on the body is always zero.   Therefore, the force  $\vec{F}$ that accelerates the test-body K with  acceleration $\vec{a}$, must be  
\begin{equation}
\vec{F}=-\vec{F}'_{inert}=\left(-\dfrac{1}{c^{2}}m\varphi_{g}\right)\vec{a}=m_{in-g0}\vec{a}
\end{equation}
The equation (4.14) is   Newton's Second Law,  for non-relativistic
	velocities,  which obviously  results from the Law of Motion.  
Therefore, the 
inertial rest mass $m_{in-g0}$ of the test-body K is  
\begin{equation}
m_{in-g0}=\left(-\dfrac{1}{c^{2}}m\varphi_{g}\right)
\end{equation}
	We will call  the inertial rest mass $ m_{in-go}$ of the test-body K,  gravitational inertial rest mass  and its momentum $\vec p_{g}$, gravitational  momentum,  because they are due to  the  gravitational potential of the rest of the Universe.
It is important to emphasize that the gravitational inertial rest mass of a body is merely a component (a coefficient) of the inertial force and thus only appears when the body is accelerated. It makes no sense when the body is moving uniformly.
So, the gravitational   inertial rest mass of a body, without internal
structure, is not an intrinsic property of the body but is proportional
to the negative of its gravitational potential energy with the rest of the Universe.

 For non-relativistic velocities the gravitational energy of the test body K   is   $ E_{g}=m_{in-g0}c^{2}=-m\varphi_{g}$
while its gravitational potential energy is $\ U_{g}=m\varphi_{g} $.  
Therefore, its  total gravitational  energy  is zero. Since  
the Universe consists of bodies such as the test body K (ignoring internal structure) the total gravitational  energy of the Universe is zero! It’s noteworthy that Richard Feynman writes in   “Lectures on Gravitation”
\cite{key-35}: 
\begin{quote}
	“Another spectacular coincidence relating the gravitational constant
	to the size of the universe comes in considering the total energy.
	The total gravitational energy of all the particles of the universe
	is something like GMM/R, where R=Tc, and T is the Hubble’s time...If now we compare this number to the
	total rest energy of the universe, $Mc^{2}$, lo and behold, we get
	the amazing result that $GM^{2}/R=Mc^{2}$, so that the total energy
	of the universe is zero. Actually, we don’t know the density nor that radius well enough to claim equality, but the
	fact that these two numbers should be of the same magnitude is a truly amazing
	coincidence...Why
	this should be so is one of the great mysteries and therefore one
	of the important question of physics. After all, what would be the
	use of studying physics if the mysteries were not the most important
	things to investigate?”
\end{quote}

If we consider that the density of matter is roughly
uniform throughout space, then the most distant matter dominates the
gravitational scalar potential, because although the influence
of matter decreases with the distance, the amount of matter goes up
as the square of the distance. Therefore, the distant
matter is of predominant importance, while local matter has only a very
small effect. Thus, it is difficult to observe any difference in the gravitational inertial rest mass with local experiments.

Consider a test body with gravitational mass $m$ and internal structure (i.e., a composite body) that is free-falling in the gravitational field of a large body which has   spherically symmetric  distribution of mass  $M$ with  $M\gg m$, in a region of the lab frame where the gravitational scalar potential from the rest of the Universe is 
$\varphi_{g}$. As we have established, the internal structure of a body in free fall does not influence its motion; only the external gravitational inertial force acts on it. Therefore, for non-relativistic velocities, applying Newton's Law of Universal Gravitation and Newton's Second Law yields the following expression for the magnitude of the body's radial acceleration
\begin{equation}
G\dfrac{Mm}{r^{2}}=\left(-\dfrac{1}{c^{2}}m\varphi_{g}\right)a
\end{equation}
where $r$ is the distance of the test body from the center of the large mass $M$. It is obvious that the gravitational mass m  of the test body is canceled
 	in equation (4.16).
 	This shows that the acceleration of a free-falling body is independent of its gravitational mass, leading to the conclusion that all bodies fall at the same rate in a gravitational field.
This is a fundamental experimental result that was tested with great accuracy with the Eötvös experiment. 
      We must emphasize that this was a free fall experiment.
It’s noteworthy that James Hartle  writes  for the Eötvös experiment \cite{key-36}:  
\begin{quote}
	“The masses are free to move in the direction perpendicular to both the fiber and the rod. Gravity is the only force acting in this 
	“twisting direction” along which the masses are effectively freely falling. Any difference between the acceleration of the two masses would cause the pendulum to twist.” 
\end{quote}
In Einstein's General Relativity, this experimental observation is explained by postulating the equivalence of gravitational and inertial mass.

 Let us now  determine the gravitational inertial rest mass of the test-body K, with gravitational mass $m$ and no internal structure for  relativistic velocities. It is well known from the Special theory of relativity that  if we wish to salvage Newton's law of momentum conservation, we must  define the gravitational momentum  $ \vec p_{g}$ of the test-body K in an inertial frame of reference S, where the test-body K  moves with velocity  $\vec u$,  as follows 
\begin{equation}
\vec p_{g}=\dfrac  {m_{in-g0}} {  \sqrt{1-{u^{2}}/{c^{2}}} }  \vec u =\gamma(u)m_{in-g0}\vec u
\end{equation}
where  the gravitational inertial rest mass $m_{in-g0} $  of the test-body K  must  be a Lorentz invariant, i.e., all observers agree on its value at any instant of test-body's history. 

For non-relativistic velocities, the gravitational inertial rest mass of the test-body  K is described by equation (4.15). Now, let's derive the equation that describes the gravitational inertial rest mass for relativistic velocities. This means we need to determine the gravitational potential energy of the entire Universe as observed by the test-body K, when it moves with velocity $ \vec {v}$ relative to the lab frame. 
The instantaneous sum of the gravitational four-potentials of all the
bodies in the Universe,  at a certain point, is also a four-vector,
the total gravitational four-potential    
\begin{equation}
A_{g}=\left(\dfrac{\varphi_{g}}{c},\vec{A}_{g}\right)
\end{equation}
and the four-velocity of the  test-body K  is: $ U=\gamma(v)(c,\vec{v}) $.

We know that the quantity we are looking for must depend on both
$U$  and  $A_{g} $, and it is a scalar.
The product $m U A_{g}$ has physical dimensions of energy and it is a scalar, because the gravitational
mass  is a  scalar, and the product of two four-vectors
is a  Lorentz invariant, i.e. a scalar.  
Evaluating the product $m U A_{g}$ in the local proper frame of the test-body  K, i.e., the K frame, where the gravitational scalar potential from the entire Universe according to equation (4.4)  is $\varphi'_{g} $, we get
\begin{equation}
m  UA_{g}=m(c,o)\left(\dfrac{\varphi_{g'}}{c},\vec{A'_{g}}\right)=m\varphi'_{g}
\end{equation}
Thus we obtain the gravitational potential energy  of the entire Universe as observed by the  test-body K,  which is the very thing we wanted   and is a Lorentz invariant.  So, for relativistic velocities, the gravitational inertial rest mass of the test-body K is 
\begin{equation}
m_{in-g0}=-\dfrac{1}{c^{2}}m  UA_{g}=-\dfrac{1}{c^{2}}m\varphi'_{g} 
\end{equation}
Substituting for $\varphi'_{g} $ from equation (4.4) into equation (4.20) we get
\begin{equation}
m_{in-g0} =-\dfrac{1}{c^{2}}m\varphi'_{g}  = \gamma(v)\left(-\dfrac{1}{c^{2}}m\varphi_{g}\right)
\end{equation}
Thus, the gravitational inertial rest mass of a body without internal structure is not an intrinsic property of the body, but is proportional to the negative of the gravitational potential energy of the entire Universe relative to the body, and is Lorentz invariant.

According to equation (4.21), the gravitational inertial rest mass of a body without internal structure is influenced by its velocity relative to the fixed stars. However, when a body has an internal structure, the inertial rest mass arising from that structure, as we will explore below, is independent of its velocity relative to the fixed stars and instead depends solely on its velocity relative to the observer. This conclusion, combined with experimental evidence showing how the inertial mass of electrons changes with relativistic velocities, leads us to conclude that electrons have an internal structure and the majority of their inertial mass is attributable to this internal composition.

\subsection{Gravitoelectric inertial rest mass of a body   without internal structure } 
\label{sec5}

Now, let's consider  a scenario where there are other electric charges near the lab frame, distributed and moving in such a way that the electric scalar potential  $\varphi_{e}$ in the lab frame is non-zero but uniform,  so that  $\vec{\nabla}\varphi_{e}=0 $    and the magnetic vector potential  $\vec{A}_{e} $    is   zero. We'll examine what happens when a test-body K, with gravitational mass $m$ and electric charge $q$ but no internal structure, is accelerated in this lab frame. 

Given that the equations of electromagnetism have the same mathematical form as those of gravitomagnetism, the test-body K  will experience both an induced gravitational and electric inertial force. In this case, we will call  the inertial rest mass of the test-body K,   gravitoelectric inertial rest mass  $m_{in-ge0}$. Using the same method we applied earlier for the gravitational inertial rest mass, we can express the gravitoelectric inertial rest mass as
 \begin{equation}
 m_{in-ge0}=  -\dfrac{1}{c^{2}}m  UA_{g} -\dfrac{1}{c^{2}}q  UA_{e} =\gamma(v)\left[-\dfrac{1}{c^{2}}(m\varphi_{g}+q\varphi_{e})\right]
 \end{equation}
This is a Lorentz invariant quantity.
 
So, when  the instantaneous velocity of  the test-body K relative to an inertial frame S is $\vec u$  and relative to the lab frame is $\vec v$,
the   gravitoelectric momentum  $\vec p_{ge}$ of the test-body K in the frame S, will be
 \begin{equation}
\vec p_{ge}=\gamma(u) \gamma(v)\left[-\dfrac{1}{c^{2}}(m\varphi_{g}+q\varphi_{e})\right]\vec u  = \gamma(u) m_{in-ge0} \vec u
 \end{equation}

In our Universe, we have various celestial bodies (stars, black holes, neutron stars, white dwarfs, planets, asteroids, comets) and the interstellar medium (dust and gas). From the lab frame perspective, we can model the observable Universe as consisting of n discrete gravitational masses and m discrete electric charges, including individual atoms or molecules in the interstellar medium.
If all these discrete bodies, including test-body K, move at non-relativistic velocities   $v_{i}$, (i.e. $v_{i} \ll c$),  relative to the lab frame, the gravitoelectric inertial rest mass of the test-body K can be approximated as
\begin{equation}
m_{in-ge0}\approx  \dfrac{1}{c^{2}}\left(\dfrac{1}{4\pi g_{0}}\stackrel[i=1]{n}{\sum}\dfrac{mm_{i}}{r_{i}}-\dfrac{1}{4\pi\varepsilon_{0}}\stackrel[i=1]{m}{\sum}\dfrac{qq_{i}}{r_{i}}\right)
\end{equation}\
where $ G= {1}/{4\pi g_{0}}$ and $ {r_{i}}$  the distances from the lab frame, as measured in the lab frame. While approximate, this equation provides a good representation of the situation.
This result demonstrates that the gravitoelectric inertial rest mass of a body depends on the distribution of matter in the Universe, including both gravitational masses and electric charges. In most cases, the electric charges tend to cancel each other out.

\subsection{Inertial rest mass  of a body with internal structure} 
\label{sec5}
Up until now, we have only considered the test-body K without internal structure. The Special Theory of Relativity, however, provides us with the means to determine the inertial rest mass of a body with internal structure, a composite body. By applying the conservation of four-momentum in an inelastic collision, where 
$n$ free-moving particles without internal structure collide to form a composite body M, the inertial
 rest mass $m_{in-0}$ of the composite  body M   is given by  
\begin{equation}
m_{in-0}=\stackrel[i=1]{i=n}{\sum}m_{in-ge0i}+   T/c^{2}+E_{field}/c^{2}
\end{equation}
where $m_{in-ge0i}$  represents the gravitoelectric inertial rest mass of each individual particle that makes up the composite body M, $T$ is the kinetic energy associated with the relative motion of all the particles, and
$E_{field}$ is the potential energy resulting from the interactions between all the particles \cite{key-37}.
Furthermore, as is well established by the Special Theory of Relativity, the inertial rest mass of the composite body  M, is a Lorentz invariant quantity  \cite{key-38}. 

Equations (4.24) and (4.25) reveal that if we substitute the gravitational mass $m$ with $im$ (an imaginary number), the inertial rest mass, which is a measurable quantity, remains a real quantity. This suggests that gravity can indeed be described by a spin-1 gauge field where the likes attract. We discussed this important topic earlier in section 2 where we demonstrated how a spin-1 gauge field theory of gravity is possible.

\subsection{Zero  gravitoelectric inertial rest mass }  
\label{sec10}
Let us now examine an important phenomenon that occurs in the inertial rest mass of a charged particle when nearby electric charges do not fully cancel each other out. Consider, using non-relativistic physics, an accelerating particle A with gravitational mass $m$, electric charge $q$  and no internal structure, placed inside a thin spherical shell of radius $ R$  with a uniformly distributed electric charge $Q$ on its surface. It is well-known that the electric scalar potential $ \varphi_{e}$ inside such a spherical shell is constant and given by: 
\begin{equation}
\varphi_{e}= \dfrac{1}{ 4\pi\varepsilon_{0}}\dfrac{Q}{R}  
\end{equation}
The  gravitoelectric inertial rest  mass of the particle  A, according to equation (4.24), is
\begin{equation}
m_{in-ge0}=\dfrac{1}{c^{2}}\dfrac{1}{4\pi g_{0}}\stackrel[i=1]{n}{\sum}\dfrac{mm_{i}}{r_{i}}-\dfrac{1}{c^{2}}\dfrac{1}{4\pi\varepsilon_{0}}\dfrac{qQ}{R}=m_{in-g0}-\dfrac{1}{c^{2}}\dfrac{1}{4\pi\varepsilon_{0}}\dfrac{qQ}{R}
\end{equation}
From equation (4.27) we can see that the gravitoelectric rest mass of the particle A can become zero if the radius of the spherical shell reaches a critical value, $R_{critical}$ given by   
\begin{equation}
m_{in-g0}=\dfrac{1}{c^{2}}\dfrac{1}{4\pi\varepsilon_{0}}\dfrac{qQ}{R_{critical}}\Longleftrightarrow   R_{critical}=\dfrac{1}{c^{2}}\dfrac{1}{4\pi\varepsilon_{0}}\dfrac{qQ}{m_{in-g0}}
\end{equation}
If we assume that particle A is an electron (with no internal structure) with an electric charge 
 $q=-1,6 \times10^{-19} C $, and that the electric charge 
 $Q$ of the spherical shell is $Q=10^{15} \times q=-1,6 \times10^{-4} C$,    then  the critical radius $R_{critical} $ is approximately 
2,81 m. If instead $Q=q$, then the $R_{critical} $ becomes approximately   
$2,81\times10^{-15}m $.
Thus, the effect of the electric scalar potential on the inertial rest mass of a charged particle can be significant, particularly in the subatomic world.
This phenomenon could be experimentally tested by measuring the inertial mass of moving electrons in a magnetic field, while the entire setup is placed inside a negatively charged spherical shell.

\section{Implications for Dark Matter and Dark Energy.}
\label{sec11}  

\subsection{Dark matter }
\label{sec9}

Stars are free-falling in the gravitational field of their galaxy, and therefore their internal structure does not affect their motion. From equations (4.24) and (4.25), it follows that the inertial rest mass of a star depends on the gravitational scalar potential of the entire Universe; in other words, the inertial rest mass of a star depends on the distribution of matter in the Universe. In the Universe, there are planets, stars, galaxies, clusters of galaxies, and so on. Therefore, the position of a star affects its inertial rest mass. In regions with higher matter density, the inertial rest mass of a star will be greater than that of an identical star in a region with lower matter density.

Moreover, since the gravitational scalar potential and the gravitational vector potential satisfy the wave equation and propagate at the speed of light, they must exhibit similar behavior to light. According to the principle of Equivalence, this means they will bend when passing near a large gravitational mass. Therefore, the gravitational potentials of the entire Universe are more concentrated in regions with higher matter density. This provides a second reason why the inertial rest mass of a star is greater in regions with higher matter density than in regions with lower matter density.

Thus, the inertial rest mass of a star near the center of a galaxy is greater than that of an identical star at the edges of that galaxy. Therefore, stars at the edges of a rotating spiral galaxy are moving faster than Newtonian physics predicts, assuming that the inertial rest mass it's the same everywhere. This phenomenon has been observed, and the inability to explain it has led to the theory of dark matter. The ideas presented above are highly likely to contribute to solving the dark matter problem.

\subsection{Dark energy }
\label{sec9}

Consider now the light emitted by an atom located on the surface of a static, spherically symmetric star with gravitational mass M. Assume that this atom, with gravitational mass $m$ and inertial  rest mass $m_{in-0}$,  is at a distance $r_{em}$ from the center of the star.  The relationship between the frequency $ f_{em}  $ of the emitted light and the frequency
$f_{\infty}$ of the light observed at infinity is given by equation \cite{key-39}
\begin{equation}
f_{\infty}=f_{em}{\sqrt{1-\dfrac{m}{m_{in-0}}  \dfrac{2GM}{c^{2} r_{em}}}}
\end{equation}
Equation (5.1) describes the well-known gravitational redshift effect, where light emitted by an atom in a gravitational field is observed at a lower frequency by an observer situated far from the gravitational source, essentially at infinity

As the Universe expands, the distances between distant galaxies and any given star increase over time. According to equation (4.24), this expansion leads to a gradual decrease in the star's inertial rest mass. Similarly, the inertial masses of atoms on the star's surface also diminish over time due to the Universe's expansion. Consequently, as equation (5.1) indicates, the light emitted by an atom on the star's surface becomes increasingly redshifted over time as a result of the expanding Universe. Thus, two identical supernovae occurring at different times in the history of the Universe will have different inertial masses. Therefore, the light emitted by the atoms of these two identical supernovae at different times in the Universe's history will exhibit different redshifts. The atoms of a younger (more recent) supernova will have smaller inertial rest mass than the atoms of an older supernova because of the Universe's expansion. Hence, the light emitted by the atoms of a younger Type Ia supernova will have a larger redshift than the light emitted by the atoms of an older Type Ia supernova. This phenomenon has been observed, and the inability to explain it has led to the theory that the Universe is expanding at an accelerating rate due to dark energy. It is very likely that the above idea can help in the  solution to this problem.

\subsection*{Conclusions }
\label{13}

In this paper, we have demonstrated that understanding the origin of inertia and the nature of inertial mass requires embracing two key ideas. First, the fundamental idea of the relativity of all kinds of motion must be accepted. Second, gravity should be described by a gravitomagnetic theory, just like the electromagnetic theory. By accepting that gravitational mass, distinct from inertial mass, is a Lorentz invariant and substituting gravitational mass m with im (an imaginary mass), gravity can be modeled as a spin-1 gauge field theory where like masses attract.
These two key ideas led us to the Equivalence Principle and the conclusion that the gravitomagnetic field influences the geometry of spacetime, which is a Finsler-Randers spacetime. However, we have shown that for a body with gravitational mass, the concept of a curved Finsler–Randers spacetime is physically equivalent to the concept of a flat Minkowski spacetime with an additional gravitomagnetic field.
By using the concept of flat Minkowski spacetime with a gravitomagnetic field, we explained the origin of inertial forces, establishing that they are real and not fictitious. Our findings indicate that external inertial forces arise as inductive effects from the entire Universe, while internal inertial forces are determined by the body's internal structure. Moreover, we have shown that during free fall in a gravitational field, the internal structure of a body does not influence its motion, explaining why all bodies fall at the same rate, even though gravitational mass is not equivalent to inertial mass.
This exploration of inertial forces has revealed that the inertial mass, a real and not imaginary quantity, is influenced by the distribution of matter throughout the Universe. The new theory is consistent with Mach's principle and provides an equation for inertial rest mass, offering significant implications for our understanding of dark matter and dark energy.

\end{document}